\documentclass{llncs}
\usepackage{graphicx} 
\usepackage{subfigure}
\usepackage{wrapfig,lipsum,booktabs}

\usepackage{amssymb}
\usepackage[fleqn]{amsmath}

\begin{document}
\title{Smart-Lock Security Re-engineered using Cryptography and Steganography}
\author{Chaitanya Bapat, Shivani Inamdar, Ganesh Baleri \and Anant V Nimkar}
\institute{Sardar Patel Institute of Technology\\ University of Mumbai, India\\
\email{chaitanya.bapat@spit.ac.in, shivani.inamdar@spit.ac.in, ganesh.baleri@spit.ac.in, anant\_nimkar@spit.ac.in}
}

\maketitle
\begin{abstract}
After the rise of E-commerce, social media and messenger bots, rapid developments have been made in the field of connecting things, gadgets, and devices, i.e, the Internet of Things (IoT). In the fast-paced lifestyle, it is very difficult to maintain multiple keys for traditional mechanical locks. Electromagnetic smart locks are a possible solution to this problem. To connect a smart lock with a key, Bluetooth Low Energy (BLE) protocol can be used. BLE protocol is vulnerable to Man-in-the-Middle (MITM) attack. Ensuring security over BLE is an ongoing challenge. This paper aims to analyze the MITM vulnerability of BLE and develop a possible solution for designing smart-locks with an increased level of security. The observation shows that the combination of Image Steganography and Cryptography helps to overcome the vulnerabilities of BLE protocol. 
\end{abstract}
%%-------------------------------------------------------------------------------------------------------
\section{Introduction}
\label{sec:intro}
The domain of Internet of Things (IoT) has shown significant capability to drastically change the technological world. IoT systems include computing and household devices, as well as sensors. It is possible to control household devices with a tap on the mobile screen, thanks to IoT. In addition, Cisco’s Internet Business Solutions Group has predicted that the number of IoT devices will be about 20.4 billion by the year 2020~\cite{c1}. IoT devices have made people's lives easier in a number of ways. Nonetheless, security experts have expresses their concerns about the threats and vulnerabilities that these devices bring along, termed as the 'Insecurity of Things'.

Mobile devices that connect to Smart Locks using the Bluetooth Low Energy(BLE) protocol are vulnerable to various security attacks like the Man-in-the-Middle(MITM) attack. BLE is a power efficient technology which is capable of transferring data between smart-phones and IoT devices. Basically, an intruder/attacker tries to impersonate a receiver and takes hold of the communication between two parties. Such an attack is called MITM attack and is found to be carried out in BLE protocol. 

When home automation and security are under consideration, locks -either mechanical or electronic- are a necessity. However, the problem associated with any physical lock is about the key handling and management. Humans tend to be forgetful and multiple keys need to be managed hence it was replaced by electromagnetic locks. However, it still didn't address the issue of remote accessibility. In the age of smart-phones and a hyper-connected world, it is essential to control locks remotely, using hand-held devices. Hence, smart-locks have been introduced to address this concern. But the issue is, despite the promise of accessibility, ease of use and comfort associated with smart-locks, security is an imminent and constant threat. So the problem is to tackle the security threats and attacks on IoT based smart-locks.

The ongoing research in the field of Internet of Things and BLE protocol relies heavily on the usage of Cryptography. The algorithm of Advanced Encryption Standard is used for encryption and decryption. However, the research has found out problems associated with cryptography algorithms like MITM attack, masquerade attack, etc. Moreover, few papers involve usage of one-time passwords for securing the communication. However, OTP generation is an intensive task and depends on network bandwidth thus suffering from latency issues.

This paper aims at investigating the working of BLE protocol and highlights the underlying architecture designed for communication using BLE protocol. In addition, it’s vulnerabilities have been studied and a synthesis of cryptographic and steganographic techniques has been implemented so as to prevent MITM attack on BLE protocol. Such a combined approach tackles the shortcomings of the individual methods of cryptography and steganography whilst preserving the advantages of each of them.

The paper is organized as follows. Section~\ref{sec:blep} focuses on the architecture of BLE. as well as the vulnerabilities existing in BLE. Section~\ref{sec:mitm} throws more light on the MITM attack  and it’s relevance in BLE protocol. Section~\ref{sec:steg}  is a review of Steganography as a possible solution to existing problems in BLE protocol. The existing solutions in the sphere of IoT devices and BLE protocol are presented in Section~\ref{sec:esol}.  In Section~\ref{sec:obs}, a combination of Steganography with Cryptography as a possible solution is proposed. The actual implementation of the system is included in Section~\ref{sec:impl} followed by discussion of the results in Section~\ref{sec:resNdis}. Ultimately, the article is concluded in Section~\ref{sec:con}.
%%-------------------------------------------------------------------------------------------------------
\section{Related Work}
\label{sec:relwork}

\subsection{Bluetooth Low Energy Protocol}
\label{sec:blep}
BLE is a wireless technology which consumes less energy and supports short range communication. This technology has can be used in various fields such as Entertainment, Health and Sports. BLE devices have easy maintenance and can work for years on coin-cell batteries~\cite{c3}. Although low-power technologies such as Zigbee, 6 LoWPAN and Z-wave have made their mark in the market, BLE has greater deployment expectations~\cite{c2}.
%%-------------------------------------------------------------------------------------------------------
\subsubsection{Security  at the Link Layer}
\label{sec:blep:sec}

Authentication and encryption is done using the Cipher Block Chaining-Message Authentication Code (CCM) algorithm and a 128-bit AES block cipher. When connection is based on encryption as well as authentication, a 4-byte Message Integrity Check (MIC) gets appended to the data channel PDU. The Payload and MIC fields are then encrypted. Authenticated data is passed over an unencrypted channel by using digital signatures. An algorithm which makes use of a  128-bit AES block cipher helps generate the signature~\cite{c2}. A counter is given as one of the inputs to this algorithm, that gives protection against various replay attacks. It is assumed that a trusted source has sent the data in case the receiver successfully verifies the signature.

For communication over BLE, pairing is an important task. Pairing in BLE is done in 3 phases.
In first phase, devices announce their input-output capabilities. Subsequently, STK (Short Term Key) is generated for secure distribution of key materials that are required for next phase. At first, both the devices agree on Temporary Key (TK). It is done using Out of Band communication, Passkey Entry or JustWorks. Based on the TK and random values generated by both the devices, STK is generated. Later, in the next phase each end-point sends to every other end-point , three 128-bit keys: Long-term key, Connection Signature Resolving Key, Identity Resolving Key. Long term key is for Link Layer Encryption and authentication. Connection resolving key performs data signing at ATT layer while Identity Resolving Key generates a private address based on the public address of the device. The STK generated in PHASE II is used for encryption while distributing these 3 keys.
In all the three phases, the message exchange is carried out by the Security Manager Protocol (SMP).
%%-------------------------------------------------------------------------------------------------------
\subsubsection{Vulnerabilities In BLE Protocol}
\label{sec:vul}
Though BLE provides modes of security, it is still prone to a number of vulnerabilities.
\begin{itemize}
\item Eavesdropping:
Although BLE consists of security modes to protect it against vulnerabilities, there are still some loopholes in the pairing phases. A BLE device is susceptible to being tracked by a third party and subsequent eavesdropping.
\item Man-in-the‐Middle Attacks: An MITM attack takes place when an intruder secretly relays and possibly alters the communication between two devices which are communicating with each other. If an attacker could somehow trick the devices into assuming that they have been disconnected from each other, then he/she could use two Bluetooth modules to act as the master and slave devices. This would thus enable packet injection and authentication attacks.
\item Denial of Service:
Denial-of-service (DoS) attacks typically flood servers, systems or networks with traffic, thereby overwhelming the victim resources. As the victim’s resources are exhausted, it becomes difficult or nearly impossible for legitimate users to use them. In DoS attacks, a server or system providing some service is attacked with a large number of requests, which results in a system crash and eventual draining of the system’s battery life. 
\end{itemize}
%%-------------------------------------------------------------------------------------------------------
\subsection{Man-In-The-Middle Attack}
\label{sec:mitm}
In order to better understand the working of MITM attacks, the paper~\cite{c4} was reviewed. MITM attack is a prominent attack in computer security, which represents a pressing concern for security experts and the academia. MITM targets the data flowing between two victims, thereby attacking the confidentiality and integrity of the data itself.
%\begin{figure}[h!]
%\centering
%\includegraphics[width=\linewidth]{MITM.png}
%\caption{MITM Exchange Methodology }
%\label{MITM label}
%\end{figure}

\begin{wrapfigure}{r}{0.6\textwidth}
\includegraphics[width=\linewidth]{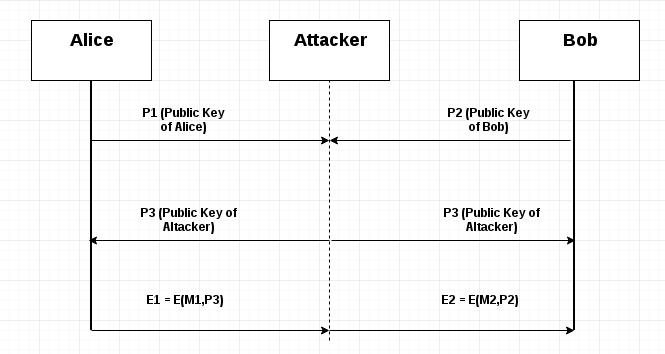}
\caption{MITM Exchange Methodology }
\label{MITM label}
\vspace{-20pt}
\end{wrapfigure}
In the MITM attack, the intruder possesses access to the communication channel between two victims, enabling him to manipulate messages flowing through the communication channel. The visualization of MITM attacks is as shown in Fig.~\ref{MITM label}. Specifically, victims try to establish a secure communication by exchanging their own public keys (P1 and P2) with each other. Attacker intercepts the public keys P1 and P2, and as a response sends its own public key (P3) to both the victims. Consequently, victim 1 encrypts its message using the attacker’s public key (P3), and sends it to victim 2 (E1). Here, as the public key used for encryption was attacker’s public key, decryption needs to be carried out using attacker’s private key. The attacker intercepts E1, and decrypts it using the corresponding private key. The attacker later encrypts some plain-text message using victim 2’s public key, and transmits it to victim 2 (message E2). 
When victim 2 is able to decrypt the messages sent by victim 1, it means that the attacker has been able to deceive both the victim parties that they are communicating over a secure channel.

MITM attack can be carried out in various communication channels such as UMTS, Long-Term Evolution (LTE), Wi-Fi, GSM, Bluetooth, and Near Field Communication (NFC). MITM attack aims to compromise:
\begin{enumerate}
\item Confidentiality- by eavesdropping on the communication.
\item Integrity- by intercepting and modifying the exchanged data .
\item Availability- by intercepting, destroying and/or modifying messages, causing one of the victims to terminate communication~\cite{c5}.
\end{enumerate}

There are minimum three ways of characterizing MITM attacks, based on:
\begin{enumerate}
\item Impersonation techniques
\item Communication channel in which the attack is executed.
\item Location of intruder and victim in the network.
\end{enumerate}
\subsection{Steganography}
\label{sec:steg}
%%-------------------------------------------------------------------------------------------------------
% \subsection{History}
% \label{sec:steg:his}
% Steganography has been in wide use ever since the Greek messengers existed. At that point in time, people in ancient Greece used the principle of steganography by concealing the wooden inscriptions using wax. This practice guaranteed confidentiality since it concealed the data. This method gained so much traction that it was resorted to again in the 1914-1918 war by the German spies. This technique of data-hiding is termed as steganography, and has become a specimen of active research in academia.
%%-------------------------------------------------------------------------------------------------------
\subsubsection{Techniques}
\label{sec:steg:tech}

Unlike cryptography, steganography does not transform the structure of the message but instead, hides it in such a way that its existence remains unidentified. There are several types of steganography, but the difference between them lies in the technique of hiding data. It is difficult to label one mechanism as the best one since each technique is chosen as per the application it is being used for. 

Hiding a message is the basis of any steganographic technique. Steganography can be classified into 2 types- technical and text. With technical steganography, confidential information can be hidden in image/audio/video files.

Text steganography, on the other hand, refers to the technique of concealing text data within a larger text. Linguistic methods are further classified into numerous categories depending on the way in which the stego-text is exploited for embedding the secret message in it. One such type of text steganography is format-based methods. These methods usually manipulate the text by changing its formatting, intentional misspelling or changing the text size. Another method is the random and statistical generation method. It can be used to prevent any comparison with the original text since based on a randomized algorithm.

%%-------------------------------------------------------------------------------------------------------
\section{Existing Solutions}
\label{sec:esol}
Application of security techniques in the field of Internet of Things is a rather new concept. However, it has been implemented in Internet Banking. It is a field where security holds prime importance and like IoT devices vulnerable to security attacks. 

The AES encryption algorithm is assumed to be the most effective for high-end security applications However, in 2011, researchers at Microsoft discovered that AES is not completely secure~\cite{c6}. Hence,alternative techniques for high-end security were studied. One such technique found was Steganography. Steganography is an approach of concealing secret information by embedding it in an image, text, audio or video. The motive of steganography is to hide the very existence of the data in any given form.

%\begin{figure}[h!]
%\centering
%\includegraphics[width=\linewidth]{Stego_Layer.png}
%\caption{Stego-layer method}
%\label{Stego-layer}
%\end{figure}

\begin{wrapfigure}{r}{0.6\textwidth}
\vspace{-28pt}
\includegraphics[width=\linewidth]{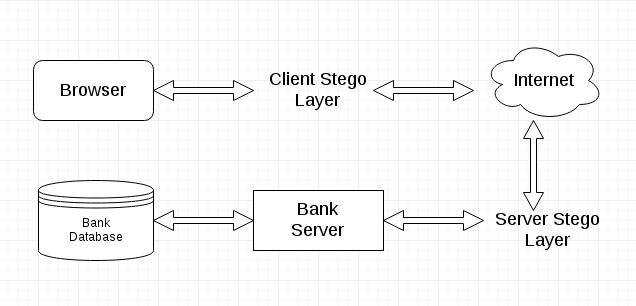}
\caption{Stego-layer method}
\vspace{-20pt}
\label{Stego-layer}
\end{wrapfigure}

Steganography has been applied in various domains. The paper~\cite{c7} proposes a ‘stego-layer’ method which provides a solution for MITM attack and Session hijacking. In the proposed method, a new ‘stego-layer’ was introduced. All the information flowing through the client or the server passes through the stego-layer. Dynamic Pattern based Image Steganography algorithm is implemented by the stego-layer for inserting and retrieving the message. Its functionality is to conceal the data to be sent between the communicating parties in an image, prior to transmission.

The authors in paper~\cite{c8} propose to improve the security of Mobile banking through Steganography. Here, the generated key determines which pixels are selected for embedding the secret message bits. The secret message bits are then planted into the selected pixels at a steady rate. However, if data bits are embedded serially in all the selected pixels, it may lead a hacker to easily hack the message.

Hiltgen, in his paper focuses on solving MITM attack by a short-time
password based on a password generating hardware token which is available from various manufacturers such as RSA Security, Active Card or VeriSign~\cite{c9}. For example, RSA’s SecureID solution consists of an LCD display and one button which enables the user to calculate the succeeding short-time password.~\cite{c10}.

Short Message Service (SMS)~\cite{c11} based One-Time Passwords (OTP) were introduced to fight phishing and other threats against authentication and authorization. The attacker's target is the possession of the password. He has various means to do so, such as a wireless interception or mobile phone Trojans. Although not very famous, the SIM Swap Attack~\cite{c12} can also be used. Through such attacks, the attacker can obtain the OTP.
AES encryption algorithm along with Steganography ensures secure and guaranteed delivery of OTP to the user. Thus, sending an OTP which is embedded in an image makes it difficult for an attacker to detect the presence of private information.

Several studies conducted on mobile malware~\cite{c13}~\cite{c14} show that the authentication credential stealing mobile malware exists in the wild.

Through all the solutions that exist, all of them fail to achieve a sure-fire way of security. May it be encryption using AES or steganography, attackers tend to find loopholes and hence pose a security threat. Usage of short-term memory passwords is limited by the hardware malfunctions and wireless interceptions. The following section proposes a solution that circumvents the given problems.
%%-------------------------------------------------------------------------------------------------------
%%-------------------------------------------------------------------------------------------------------
\section{Proposed Solution}
\label{sec:obs}
The security and privacy of any information traveling across a channel that promotes open communication results into a major problem. Hence, in order to prevent unauthenticated and unwarranted access and usage, confidentiality and integrity is needed. Of the many methods available, steganography and cryptography are two of the most used ones. The first one hides the sheer existence of the information while the second one twists the data itself~\cite{c14}. 

The data is transformed into another incomprehensible format which is then sent over the network, in case of cryptography. However, in case of steganography, stego-files such as image, text, audio, video is used as a platform for embedding the message. Later, the stego-file is transferred over the communication channel. This paper is based on harnessing the advantages of both the methods - steganography and cryptography which will facilitate an increase in the level of security.

\subsection{Workflow Design}
\label{sec:impl:wd}
%\begin{figure}[h!]
%\centering
%\includegraphics[width=\linewidth]{Workflow.jpg}
%\caption{Workflow diagram}
%\label{Workflow diagram}
%\end{figure}

\begin{wrapfigure}{r}{0.6\textwidth}
\vspace{-20pt}
\includegraphics[width=\linewidth]{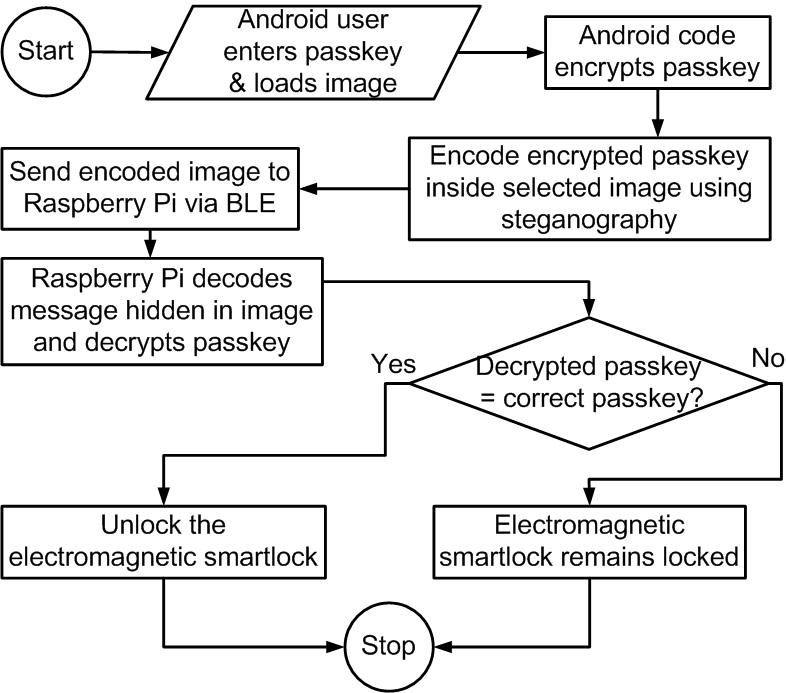}
\caption{Workflow diagram}
\label{Workflow diagram}
\vspace{-10pt}
\end{wrapfigure}

Once the system was designed, the next step in the implementation stage was to design the workflow. As depicted in Fig.~\ref{Workflow diagram} below, the techniques of Cryptography and Steganography are used hand in hand to provide security to smart-lock. User first enters the passkey via the Android smart-phone application. Later, the image is selected in which the passkey would be embedded. Using AES encryption, the passkey is first encrypted and then the encrypted cipher-text is encoded in the image. All this happens at the client-side (Android smart-phone application). Client-server architecture is utilized where server is the Raspberry Pi. Image is sent over the Bluetooth 4.0 (BLE) protocol. From the received image, cipher-text is then decoded. It is then decrypted to get the passkey entered by the user. The algorithm checks for valid passkey and accordingly takes the decision whether to open the lock or not.

In the cryptographic method, once a third party attacker or an intruder gains access to the secret key, the data gets revealed. In case of steganography, the presence of message itself gets concealed but the form of the message is not changed. As a result, the moment the attacker understands the existence of a concealed data in whichever stego-file, the message again gets revealed.

If a combination of both the methods is used, security gets enhanced considerably as both steganalysis and cryptanalysis would be needed to be carried out in synchronization so as to identify the location of original information and the actual content itself. Combination of such techniques in the domain of security is a relatively new direction. However, one can search for similar works in the literature. Primarily, this work can be found in the paper ~\cite{c15}. A system which enhances the least significant bit (LSB) method has been proposed by the authors. 

In the domain of integrating steganography and cryptography, the paper ~\cite{c16} lends some real insight. Here, the key that is important for deciphering the original message is also implanted in the stego-file.

%%-------------------------------------------------------------------------------------------------------
\section{Implementation}
\label{sec:impl}
%%-------------------------------------------------------------------------------------------------------
\subsection{System Design}
\label{sec:impl:sd}
%%-------------------------------------------------------------------------------------------------------

\begin{wrapfigure}{r}{0.5\textwidth}
\vspace{-28pt}
\includegraphics[width=\linewidth]{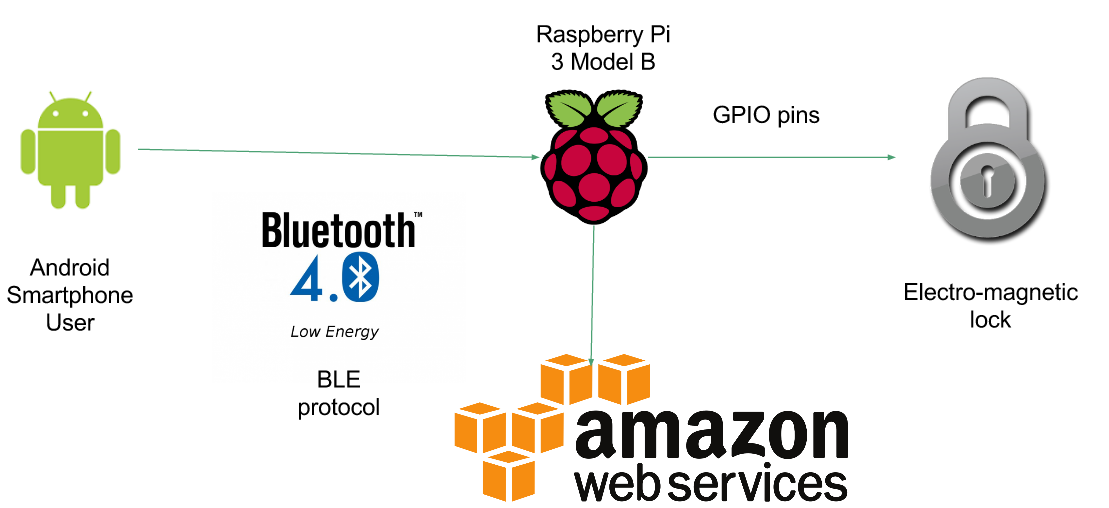}
\caption{System Design }
\label{System Design}
\vspace{-20pt}
\end{wrapfigure}
In order to create a remote-controlled system for accessing the electro-magnetic lock, the system was first designed, as shown in Fig.~\ref{System Design}. It consisted of 4 main components - Android smart-phone, Raspberry Pi, Electromagnetic lock and Server. The Bluetooth and WiFi modules in Android are fairly robust with good documentation support. Of the multiple versions of Raspberry Pi, the latest Pi 3 Model B has inbuilt WiFi and Bluetooth 4.0 (BLE). Server would be needed to log the system usage so as to provide future scope for performing analytics and understand usage patterns and statistics. 
%%-------------------------------------------------------------------------------------------------------
%%-------------------------------------------------------------------------------------------------------
\subsection{Circuit Design}
\label{sec:impl:sd}
\begin{figure}[h!]
\centering
\includegraphics[width=0.8\linewidth]{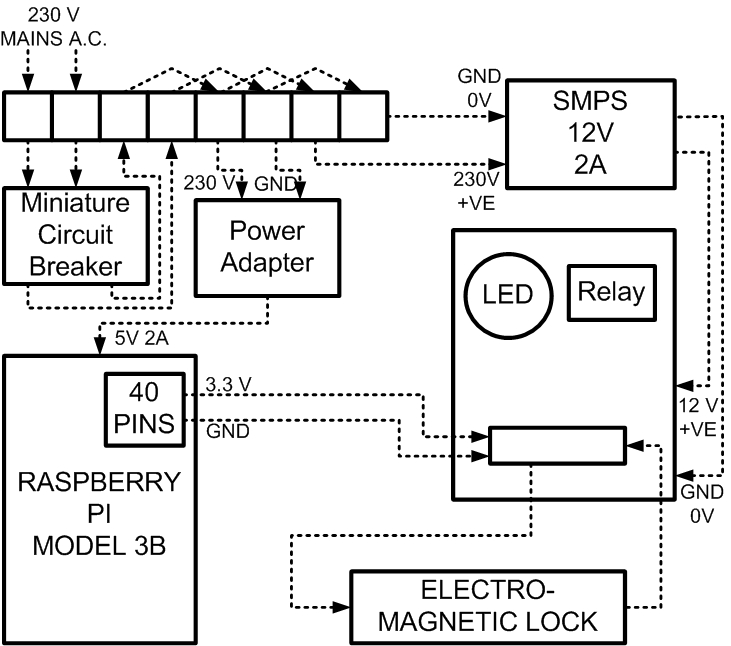}
\caption{Circuit diagram}
\label{Circuit diagram}
\end{figure}

The hardware requirements involved designing the Circuit with main components being smart-lock i.e. an electro-magnetic lock with Raspberry Pi 3 Model B. As shown in Fig.~\ref{Circuit diagram}, the mains 230V alternating current is fed to the miniature circuit breaker (MCB) which breaks the circuit during power failure or short-circuit. It prevents any damage to the internal circuit components. Power adapter facilitates conversion of 230 V power supply to 5V as needed by Raspberry Pi. Raspberry Pi provides 3.3V with respect to its ground as an output to general-purpose input output(GPIO) pins. A relay acts as a switch for accessing the lock. However, the relay circuit works on 12V supply provided by SMPS. The circuit is completed by connecting electromagnetic lock in series with the relay and joining the grounds of Raspberry Pi and electromagnetic lock.

%%-------------------------------------------------------------------------------------------------------
\section{Results and Discussion}
\label{sec:resNdis}
\begin{figure}
\centering
\begin{minipage}{.5\textwidth}
\centering
\includegraphics[width=\linewidth]{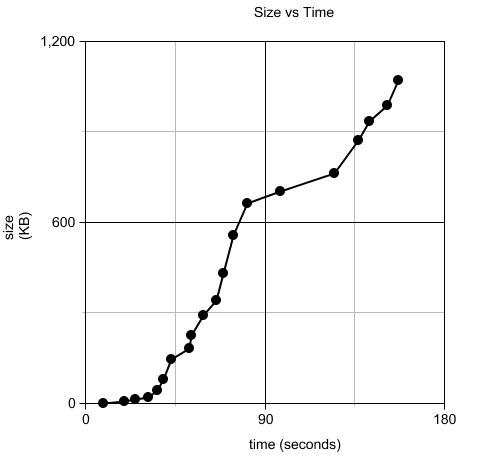}
\caption{Image size vs Total Time taken}
\label{Image size vs Total Time taken}
\end{minipage}%
\begin{minipage}{.5\textwidth}
\centering
\includegraphics[width=\linewidth]{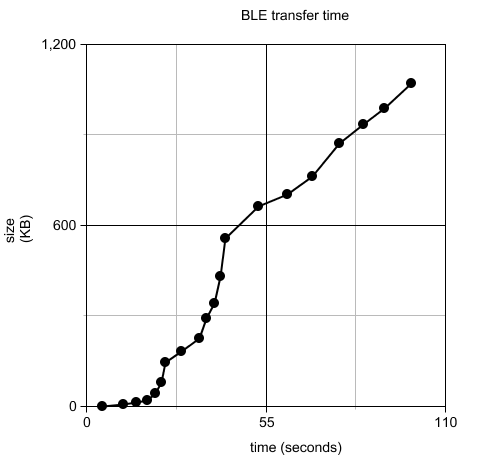}
\caption{Image size vs BLE Transfer time }
\label{Image size vs BLE Transfer time}
\end{minipage}
\end{figure}

The graph~\ref{Image size vs Total Time taken} shows the relationship between image size and total time taken. Total time taken is the time needed for image to be encoded, encrypted, sent over BLE protocol, received, decoded and decrypted. Thus, lesser the image file size, faster the communication and processing.

To find out the efficiency of the BLE protocol, the time needed only for transferring image was first tracked. The BLE transfer time takes into account the time needed to send the image from Android smart-phone to the Raspberry Pi 3 over BLE protocol. It was found that there exists a fairly linear relationship between the image size and BLE transfer time. Hence lesser image sizes would be transferred faster, as per Fig.~\ref{Image size vs BLE Transfer time}.

\begin{table*}[!h]
\caption{Dimension table}
\centering  
%\resizebox{0.99\textwidth}{!}{
\begin{tabular}{|c|c|c|} \hline
Dimensions & Image Size (kb) & Time (sec) \\ \hline
225 * 400 & 6.97 & 19.8 \\ \hline
225 * 400 & 21.85 & 22.85 \\ \hline
720 * 1280 & 43 & 36 \\ \hline
720 * 1280 & 79.7 & 36.01 \\ \hline
720 * 1280 & 224 & 52.27 \\ \hline
720 * 1280 & 557 & 64 \\ \hline
720 * 1280 & 1070 & 120.7 \\ \hline
1200 * 1200 & 1100 & 137 \\ \hline
\end{tabular}
%}
\label{Dimension table}
\end{table*}

\begin{figure}
\centering
\subfigure[Preprocessing]{\includegraphics[width = 0.4\linewidth]{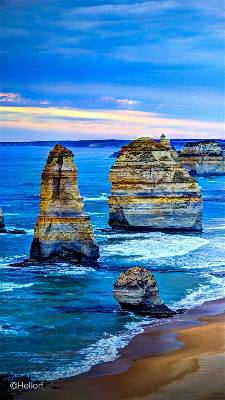}\label{Preprocessing wallpaper}}\    
\subfigure[Post-processing]{\includegraphics[width = 0.4\linewidth]{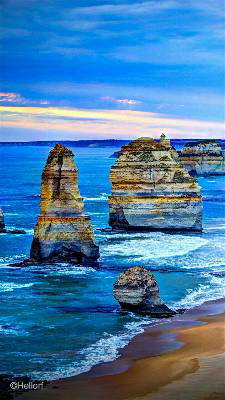}\label{Post-processing wallpaper}}
\caption{Wallpaper}
\label{Wallpaper}
\end{figure}

\begin{figure}
\centering
\subfigure[Preprocessing]{\includegraphics[width = 0.4\linewidth]{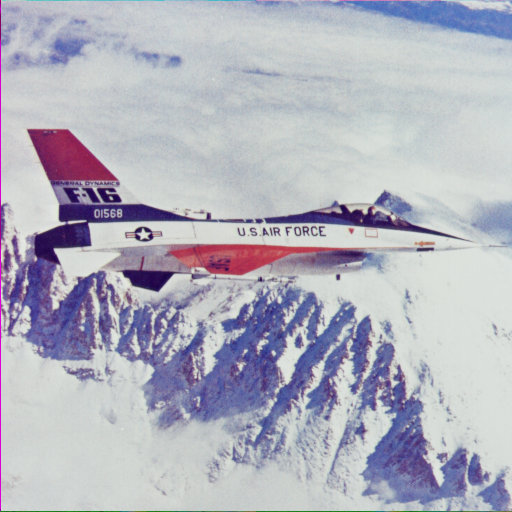}\label{Preprocessing airplane}}\    
\subfigure[Post-processing]{\includegraphics[width = 0.4\linewidth]{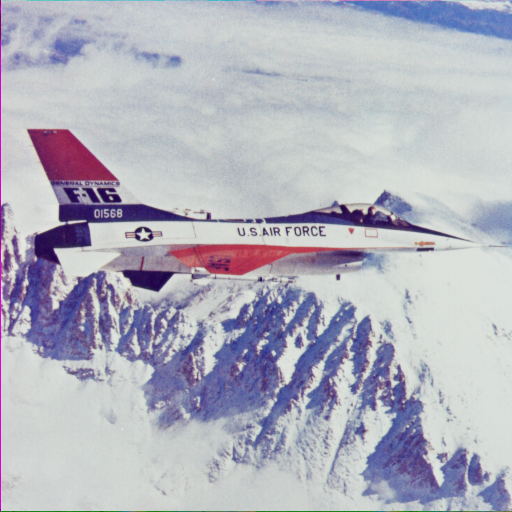}\label{Post-processing airplane}}
\caption{Airplane}
\label{Airplane}
\end{figure}

Table~\ref{Dimension table} summarizes the relationship between the image size, its dimensions and the time needed. It shows direct relationship between the image size and the time needed for processing the image.

Fig.~\ref{Wallpaper} and Fig.~\ref{Airplane} show the difficulty an attacker will experience to find out differences between the 2 images. With absolutely no visual differences in the pre-processing and post-processing image, it satisfies the requirement of providing an additional layer of security to the existing system in IoT devices.
%%-------------------------------------------------------------------------------------------------------
\section{Conclusion}
\label{sec:con}
This paper is an effort to review existential security threats in the sphere of IoT, vulnerabilities of BLE protocol and related work around MITM attacks. Having studied the BLE protocol, various issues were found including the possibility of MITM attack. Although existing solutions involve SMS One-Time-Password, Cryptography, Steganography, still few vulnerabilities persist. According to the study of these techniques, a combination of both (Cryptography and Steganography) ensures elimination of the disadvantages of the individual methods while retention of the advantages that these principles possess. An implementation of such a methodology can possibly aid research in the field of Security in IoT and fortify the future of BLE enabled IoT devices.

\bibliographystyle{splncs}

%\bibliographystyle{splncs}
%\bibliography{./FrameworkVRM}
\end{document}